# Atomic-Scale Interface Engineering of Majorana Edge Modes in a 2D Magnet-Superconductor Hybrid System

Alexandra Palacio-Morales[1,†], Eric Mascot[2], Sagen Cocklin[2], Howon Kim[1], Stephan Rachel[3], Dirk K. Morr[2,*] and Roland Wiesendanger[1,*]

[1]Department of Physics, University of Hamburg, D-20355 Hamburg, Germany

[2]Department of Physics, University of Illinois at Chicago, 845 W. Taylor St, m/c 273, USA

[3]School of Physics, University of Melbourne, Parkville, VIC 3010, Australia

†Present address: Sorbonne Université, UPMC Université Paris 06, CNRS-UMR 7588, Institut des NanoSciences de Paris, 4 Place Jussieu, 75005 Paris, France

*Correspondence to: wiesendanger@physnet.uni-hamburg.de and dkmorr@uic.edu

Topological superconductors are predicted to harbor exotic boundary states - Majorana zero-energy modes - whose non-Abelian braiding statistics present a new paradigm for the realization of topological quantum computing. Using low-temperature scanning tunneling spectroscopy (STS), we here report on the direct real-space visualization of chiral Majorana edge states in a monolayer topological superconductor, a prototypical magnet-superconductor hybrid system comprised of nano-scale Fe islands of monoatomic height on a Re(0001)-O(2×1) surface. In particular, we demonstrate that interface engineering by an atomically thin oxide layer is crucial for driving the hybrid system into a topologically non-trivial state as confirmed by theoretical calculations of the topological invariant, the Chern number.

Majorana zero-energy modes in topological superconductors are currently attracting great scientific interest because of their potential application in topological quantum information processing based on their non-Abelian quantum exchange statistics [1–7]. Recently, a promising new route to the creation of topological superconductors has been opened in one-dimensional nano-scale hybrid systems. The reported observation of zero-energy Majorana bound states at the ends of one-dimensional (1D) Rashba-nanowire heterostructures [8–10] and of chains of magnetic adatoms on the surface of s-wave superconductors [11–14] has provided the proof of concept for the creation of these exotic quasiparticles in condensed matter systems. Similarly, it was proposed that two-dimensional (2D) topological superconductors can be created by placing islands of magnetic adatoms on the surface of s-wave superconductors [15–19] (see Fig. 1**A**). In such systems, chiral Majorana modes – the variant of Majorana zero-energy states in 2D – are predicted to be localized near the edge of the island (see Fig. 1**A**), and to form a dispersing, one-dimensional mode along the edge that traverses the superconducting gap (Fig. 1**B**). This, in turn, opens new possibilities for the experimental identification of Majorana modes, as they are expected to exhibit a characteristic energy and spatial structure and to be robust against edge disorder, two properties that can be uniquely explored and visualized via scanning tunneling spectroscopy. Moreover, the ability to use atomic manipulation techniques [12,20] to create islands of particular shapes, to control their disorder at the atomic scale, and the possibility to spatially move Majorana states through the creation of superconducting vortices [3,5,17], opens unprecedented opportunities for the *quantum design* of Majorana modes in such heterostructures.

To demonstrate the creation of topological superconductivity and the ensuing chiral Majorana modes through interface engineering in 2D magnet-superconductor hybrid structure, we have grown nano-scale Fe islands of monoatomic height on the (2×1) oxygen-reconstructed (0001) surface of the s-wave superconductor Re under ultra-high vacuum (UHV) conditions. The insertion of an atomically thin oxide separation layer between the magnetic Fe island and the superconducting Re surface is shown to be crucial for the emergence of a topologically non-trivial superconducting state, which is absent when the Fe island is deposited directly on the surface of the superconducting Re substrate. The spatially- and energy-resolved differential tunneling conductance, $dI/dU$, was measured in situ using a low-temperature scanning tunneling microscope (STM) with a superconducting Nb tip to improve the energy resolution (see Supplemental Information (SI) Materials and Methods).

A constant-current STM image of the Fe/Re(0001)-O(2×1) hybrid system with a single nano-scale Fe island is shown in Fig. 1**C**. In general, it is not possible to identify the relative positions of the Fe- and O-atoms with respect to the Re surface atoms from STM images. However, DFT studies [21] suggest that the O-atoms are located above the hcp hollow sites of the Re(0001) surface, forming a p(2×1)structure, as confirmed by topographic surface profiles from the STM image (see SI Sec. 1). Moreover, our theoretical analysis discussed below suggests that the Fe atoms are located directly above the surface Re atoms, thereby continuing the ABAB stacking of the Re bulk lattice with the intermediate O atoms located above the Re hcp hollow sites. The resulting atomic structure of the hybrid system is displayed in Fig. 1**D.** The raw $dI/dU$ measured above the O(2×1) surface with a superconducting tip is shown in Fig. 1**E** (see SI Sec. 1). To account for the energy dependence of the density of states in the tip, we deconvolute the raw $dI/dU$ using standard methods (see SI Sec. 2), yielding the deconvoluted $dI/dU$ shown in Fig. 1**F**. The energies of the coherence peaks reveal a superconducting gap of $\Delta_{O(2\times1)} \approx 280$ μeV which is slightly lower than that measured on the pure Re(0001) surface ($\Delta_{Re} \approx 330$ μeV). Spectroscopic measurements on different structural domains of the O(2×1) layer indicate a uniform spatial distribution of the superconducting gap.

A necessary requirement for the emergence of topological superconductivity is that the Fe islands couple magnetically to the superconducting Re surface. The observation in $dI/dU$ of a Yu-Shiba-Rusinov (YSR) in-gap state [22–26] near an isolated magnetic Fe adatom located on top of the O(2×1) surface demonstrates the existence of the magnetic coupling to the superconducting Re surface, despite the presence of an intermediate oxide layer (see SI Sec. 1). Moreover, the presence of the nanoscale Fe island gives rise to the formation of a band of YSR states near the O(2×1) gap edge [17]. As a result, $dI/dU$ measured in the middle of the Fe island reveals a smaller superconducting gap $\Delta_{Fe} \approx 240$ μeV than that of the O(2×1) layer (see Fig. 1**F**).

A hallmark of 2D topological superconductors is the existence of dispersive, in-gap Majorana modes that are spatially located along the edges of the system [15,17,27] (Figs. 1**A** and **B**). To visualize the existence of such modes, we present in Figs. 2**A-F** (Figs. 2**G-L**) the raw (deconvoluted) spatially resolved differential tunneling conductance, in and around the nano-scale Fe island shown in Fig. 1**C**, with increasing energy, from $E_F$ up to the energy of the coherence peak at $\Delta_{Fe}$. At $E_F$ (Fig. 2**G**), the $dI/dU$ map exhibits a large intensity along the edge of the Fe island (confined to within a distance of 5 nm from the edge), clearly indicating the existence of an in-gap edge mode expected for a topological superconductor. With increasing energy (Fig. 2**H** and **I**) the edge mode begins to extend further into the island, consistent with an increase in the mode's localization length, $\lambda(E)$ (Fig. 1**A**) (see discussion below). Note that the $dI/dU$ measured inside the Fe island and along the edge are of similar intensity already for energies below $\Delta_{\text{Fe}}$, when the localization length becomes comparable to the size of the Fe island (Fig. 2**J**, $E \approx \pm 140$ μeV, see discussion below). Increasing the energy even further reverses the intensity pattern, such that at the energy of the coherence peaks $\Delta_{\text{Fe}} \approx \pm 240$ μeV (Fig. 2**L**) the $dI/dU$ intensity at the edge is smaller than that in the island's center.

To theoretically understand the origin of the combined spatial and energy distribution of the experimentally observed differential tunneling conductance, we consider the Hamiltonian $H = H_{Re} + H_{Fe} + H_{FeRe}$ (for details and parameters, see SI Sec. 3), where $H_{Re}$ ($H_{Fe}$) describes the real-space intra- and inter-orbital hopping elements, the spin-orbit coupling, and superconducting pairing in the Re 5d-orbitals (Fe 3d-orbitals), and $H_{FeRe}$ represents the electronic hopping/hybridization between the Re surface and the Fe island. As the magnetic structure of the Fe island cannot be deduced from the experimental STS data obtained with a superconducting Nb-tip, we consider two magnetic structures for the Fe magnetic moments, which are typically observed on surfaces: a ferromagnetic, out-of-plane alignment [28] as well as a 120° Néel-ordered in-plane structure [29]. To directly compare the theoretically computed local density of states (LDOS) with the experimental results, we consider an Fe island with the same lattice structure and number of atoms as studied experimentally that is located on a Re(0001)-O(2×1) surface. We model the influence of the intermediate O(2×1) layer through a modification of the hybridization described by $H_{FeRe}$.

For a ferromagnetic structure, the theoretically computed spatial and energy dependence of the LDOS (Figs. 2**M-R**) agrees well with that of the deconvoluted experimental differential tunneling conductance (Figs. 2**G-L**). Similarly, the theoretically computed LDOS inside the Fe island and of the bare Re(0001)-O(2×1) surface also shows a reduction of the superconducting gap in the former (see SI Sec. 3), in agreement with the experimental findings (Fig. 1**F**). Moreover, computing the topological invariant [30] (see SI Sec. 3) for the parameters used in Figs. 2**M-R**, we obtain a Chern number $C = 20$. These results taken together strongly suggest that the edge modes shown in Fig. 2 are chiral Majorana modes arising from an underlying topological superconducting state in the Fe/Re(0001)-O(2×1) hybrid system. Note that uncertainties in the band parameters, and in particular in the hybridization strength, might affect the actual value of the Chern number, but will not result in a topologically trivial phase (see discussion below). Similarly, we find that the system is a topological superconductor for a 120° Néel-ordered in-plane structure [29] of the Fe moments (see SI Sec. 4). Since the energy and spatial dependences of the edge modes similar to those shown in Figs. 2**M-R** are also found in generic models of topological superconductors (see SI Sec. 5) [17], they should be considered a universal feature of topological superconductivity.

A further important signature of Majorana modes is that they are topologically protected against edge disorder according to the bulk-boundary correspondence [27]. Indeed, edge modes that possess a trivial, non-topological origin can easily be moved away from the Fermi energy, or even be destroyed by disorder. However, the experimentally studied Fe island does not only possess a spatial symmetry different from the underlying Re(0001)-O(2×1) surface, but also exhibits a large degree of disorder along its edges (Fig. 1**C**). The fact that despite this disorder, edge modes are observed at $E_F$, further supports our conclusion that these modes are topologically protected chiral Majorana modes.

To further elucidate the properties of the Majorana edge modes, we present in Fig. 3 the deconvoluted differential tunneling conductance (Fig. 3**A**) and the theoretically calculated LDOS (Fig. 3**B**) for increasing energy along a cut through the Fe island, as shown in the insets (the surface profile of the island is shown in the lower panels of Fig.3**A** and **B**). Both quantities agree quantitatively and decay as expected exponentially into the island with a localization length $\lambda(E)$ (as sketched in Fig. 1**A**) that increases with increasing energy. The maxima in $dI/dU$ and the LDOS are located right at the edge of the island for all energies, as expected from generic models of topological superconductors with a dominant s-wave order parameter [18] (see SI Sec. 5). This very good agreement between the topographically determined edge of the Fe island and the maximum in $dI/dU$ therefore provides additional evidence for the topological nature of the edge modes. Note that the recently reported spatial splitting of the edge mode [18] can only be understood if one assumes a topological superconductor with a predominant p-wave order parameter [18], or if the edge mode starts to hybridize with bulk states [19].

The question naturally arises to what extent the observed topological superconducting phase of the Fe/Re(0001)-O(2×1) hybrid system is robust against variations in the strength of parameters, such as the hybridization between the Re surface and the Fe island, $V_{FeRe}$, which is mediated by the O(2×1) oxide layer, or the chemical potentials, $\mu_{Fe}$ or $\mu_{Re}$. To investigate this question, we present in Fig. 3**C** the topological phase diagram of the system, as a function of $V_{FeRe}$ and $\mu_{Fe}$ (the Chern number was calculated as described in SI Sec. 3, and the parameters for the system considered above are indicated by a yellow circle). The phase diagram clearly reveals an abundance of topological superconducting phases in close proximity to each other that are characterized by different Chern numbers. While the uncertainty in the electronic band parameters we have used to describe the Fe/Re(0001)-O(2×1) system can result in an uncertainty of the actual Chern number in the experimentally realized system, the fact that the system resides in a topological phase is robust. This substantiates our conclusion that the observed edge modes directly reflect the topological nature of the magnet-superconductor hybrid system.

To demonstrate that interface engineering using an atomically thin oxide layer is crucial for the emergence of topological superconductivity in the Fe/Re(0001)-O(2×1) hybrid system, we contrast the above results with those obtained when the Fe layer is directly deposited on the Re(0001) surface without an intermediate oxide layer [29] (see Fig. 4). In this case, with the Fe atoms being located directly above the Re(0001) hcp hollow sites (Fig. 4**H**), $dI/dU$ measured on the Fe islands exhibits no signature of edge modes in the raw data (Fig. 4**E-G**) as well as in the deconvoluted data (Fig. 4**I-K**); rather it reflects delocalized, bulk-like excitations throughout the entire Fe island at $E_F$. Our theoretical analysis of this system reveals (using the same set of parameters as for Fig. 2**M-R**) that the change in the spatial location of the Fe atoms with respect to the Re(0001) surface (arising from the missing intermediate oxide layer) renders the electronic structure of this hybrid system gapless. As a result, the LDOS at $E_F$ is non-zero, leading to the delocalized excitations observed experimentally. Moreover, the absence of a gap implies that the system is in a trivial, non-topological phase, explaining the absence of edge modes, as observed experimentally. Thus the change in the relative position of the Fe and Re atoms induced by the presence of the oxide layer is crucial for realizing topological superconductivity in the hybrid system.

In conclusion, we reported the engineering of topological superconductivity and direct visualization of theoretically predicted Majorana edge modes in a nano-scale Fe island located on a Re(0001)-O(2×1) substrate. The combination of spatially resolved spectroscopy and topographic information allowed us to not only correlate for the first time the energy and spatial dependence of the observed in-gap edge modes with the physical edge of the Fe island, but also to demonstrate the robustness of these edge modes against edge disorder [no topographic information was available in a previous study using Pb/Co/Si(111) heterostructures [18]]. Both of these experimental observations, which are in very good agreement with our theoretical calculations, represent hallmarks of the modes' topological nature. Our theoretical studies also demonstrate that the emergence of topological superconductivity in Fe/Re(0001)-O(2×1) does not require any fine-tuning of parameters, but is expected over a wide range of band parameters and magnetic structures. Moreover, we demonstrated that the emergence of topological superconductivity in such a hybrid magnet-superconductor system is only made possible through interface engineering using an atomically thin separation layer. The direct real space visualization of chiral Majorana edge modes demonstrated here, in combination with STM-based single-atom manipulation techniques [12,20], opens unprecedented opportunities to realize topological phases in artificial 2D magnetic adatom arrays on elemental superconducting substrates providing fascinating building blocks for future topological quantum computer architectures.

**Acknowledgments:**
We would like to thank H. Jeschke and M. Vojta for stimulating discussions.
**Funding:** This work was supported by the European Research Council Advanced Grant ASTONISH (project no. 338802); the Alexander von Humboldt foundation; the U.S. Department of Energy, Office of Science, Basic Energy Sciences, under Award No. DE-FG02-05ER46225.

**Author contributions:** A.P.M. and R.W. conceived and designed the experiments. A.P.M. and H.K. carried out the STM/STS experiments, processed and analyzed the data. E.M., S.C., S.R. and D.K.M. performed the theoretical modelling. R.W. and D.K.M. supervised the project. All authors discussed the results and contributed to the manuscript.

**Competing interests:** Authors declare no competing interests.

**Data and materials availability:** The authors declare that the main data supporting the findings of this study are available within the article and the supplementary materials. Extra data are available from the corresponding authors upon reasonable request.

**Supplementary Materials:**

Materials and Methods

Supplementary Text

Figs. S1 to S5

Tables S1

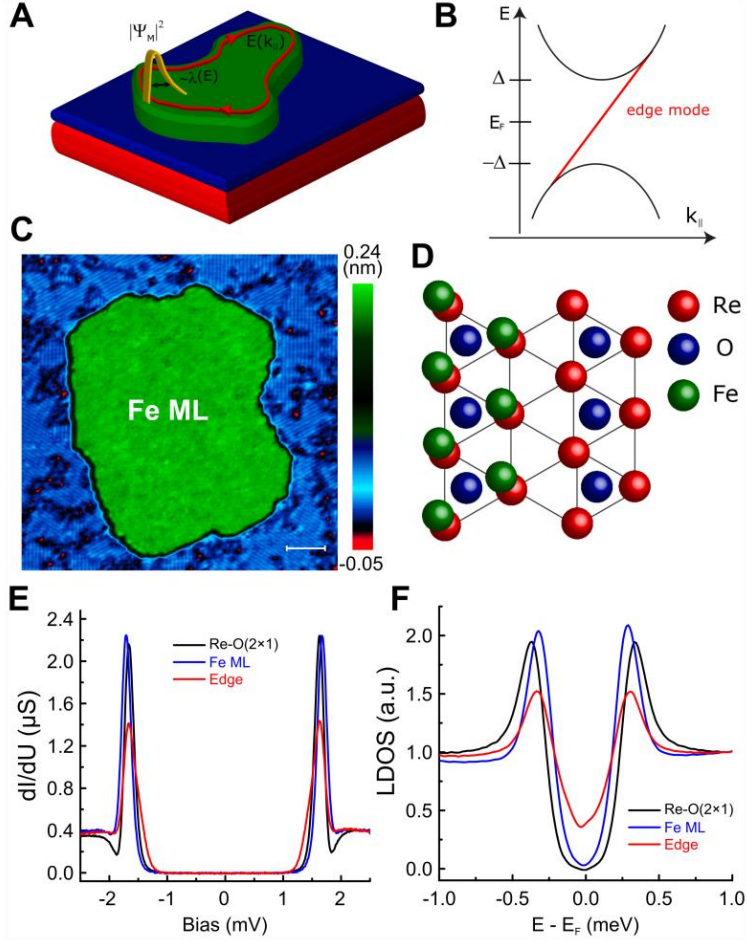

**Fig. 1. Chiral Majorana edge modes and the structure of the nano-scale magnet-superconductor hybrid system Fe/Re(0001)-O(2×1).** (**A**) Schematic picture of the Fe/Re(0001)-O(2×1) hybrid system, indicating the spatial structure of the Majorana edge modes and their spatial decay into the center of the island. Here, $k_\parallel$ is the mode's momentum parallel to the edge. (**B**) Schematic picture of the edge mode's dispersion, traversing the superconducting bulk gap. (**C**) Constant-current STM image of a small Fe island located on the O(2×1) reconstructed surface of Re(0001) (scale bar: 5 nm). (**D**) Atomic model of the hybrid system, with the Re(0001) substrate (red spheres), p(2×1) oxide layer (blue spheres representing O atoms) and Fe adatoms (green spheres). O atoms are located above the Re hcp hollow sites and Fe adatoms are located above the Re atoms. Lattice constant of the Re(0001) surface: $\mathbf{a}_{Re}$ = 0.274 nm. (**E**) Experimentally measured $dI/dU$ spectrum on the O(2×1) surface (black line), in the center (blue line) and at the edge (red line) of the Fe island. As all STS measurements were performed with a superconducting Nb tip whose gap size is $\Delta_{tip}^{Nb}$= 1.41 meV, the coherence peaks are located at $\Delta_{tip}^{Nb} + \Delta_{O(2\times 1)}$ or $\Delta_{tip}^{Nb} + \Delta_{Fe}$, respectively. This yields a measured superconducting gap of $\Delta_{O(2\times 1)} \approx 280$ μeV in the O(2×1) layer, and of $\Delta_{Fe} \approx 240$ μeV in the center of the Fe island. (**F**) Deconvoluted $dI/dU$ spectrum on the O(2×1) surface (black line), in the center (blue line) and at the edge (red line) of the Fe island. Parameters for constant-current STM images and stabilization conditions for STS measurements: $U = 2.5$ mV, $I = 1.0$ nA, $T = 360$ mK.

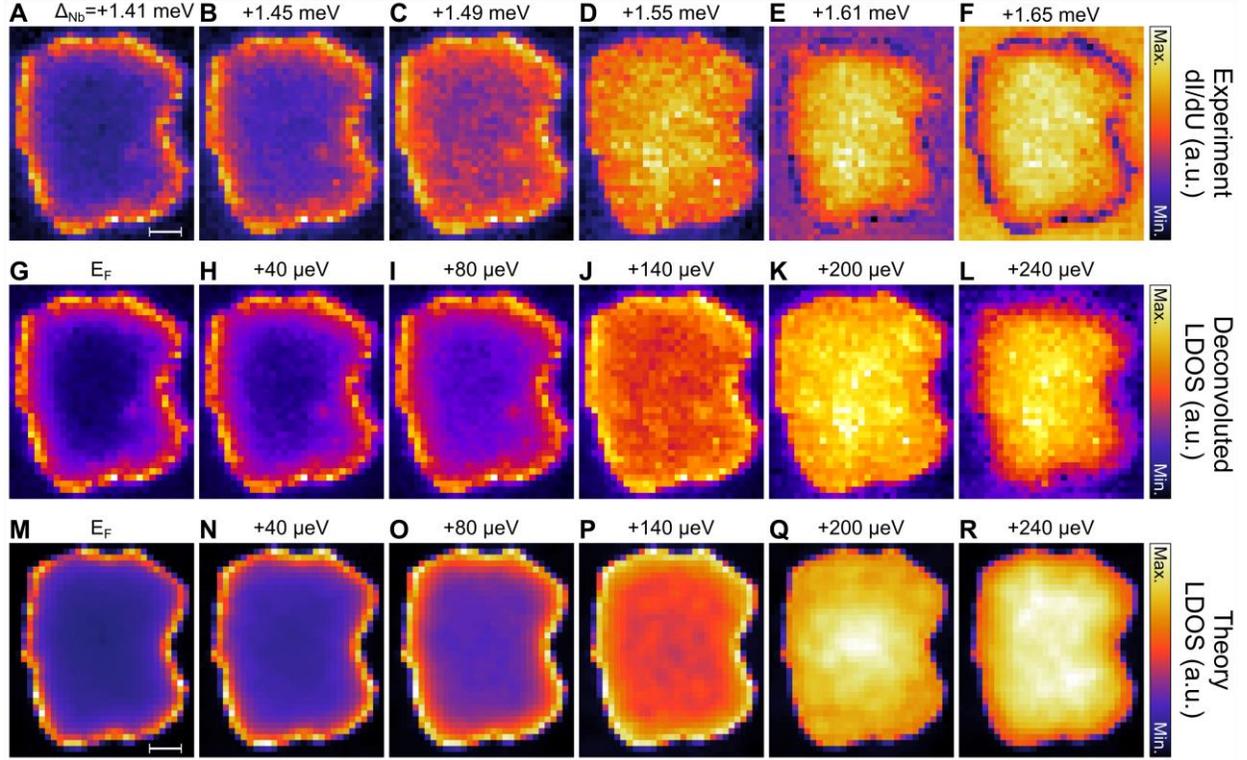

**Fig. 2. Evolution of the chiral edge states for the hybrid system Fe/Re(0001)-O(2×1) with increasing energy**. (**A-F**) Experimentally measured differential tunneling conductance maps and corresponding deconvoluted data sets (**G-L**) for the Fe island shown in Fig. 1**C** from the Fermi level $E_F = \Delta_{\text{tip}}^{\text{Nb}}$ to $\Delta_{\text{Fe}} = 240$ μeV above $E_F$. At low energies (**A-C** and **G-I**), the $dI/dU$ map reveals states that are localized along the edges of the Fe island (edge modes). Once the localization length of the edge modes becomes of the size of the island (**D** and **J**), the $dI/dU$ in the center of the island is comparable to that along the edges. (**M-R**) Theoretically computed spatially resolved local density of states for the same energies as in **G-L**. The theoretical data have been spatially convoluted to reproduce the experimental spatial resolution (see SI Sec. 3). STS measurement conditions: $U = 2.5$ mV, $I = 1.0$ nA, $T = 360$ mK. The intensity scale is adjusted for each figure separately. Scale bars in **A** and **M**: 5 nm.
10

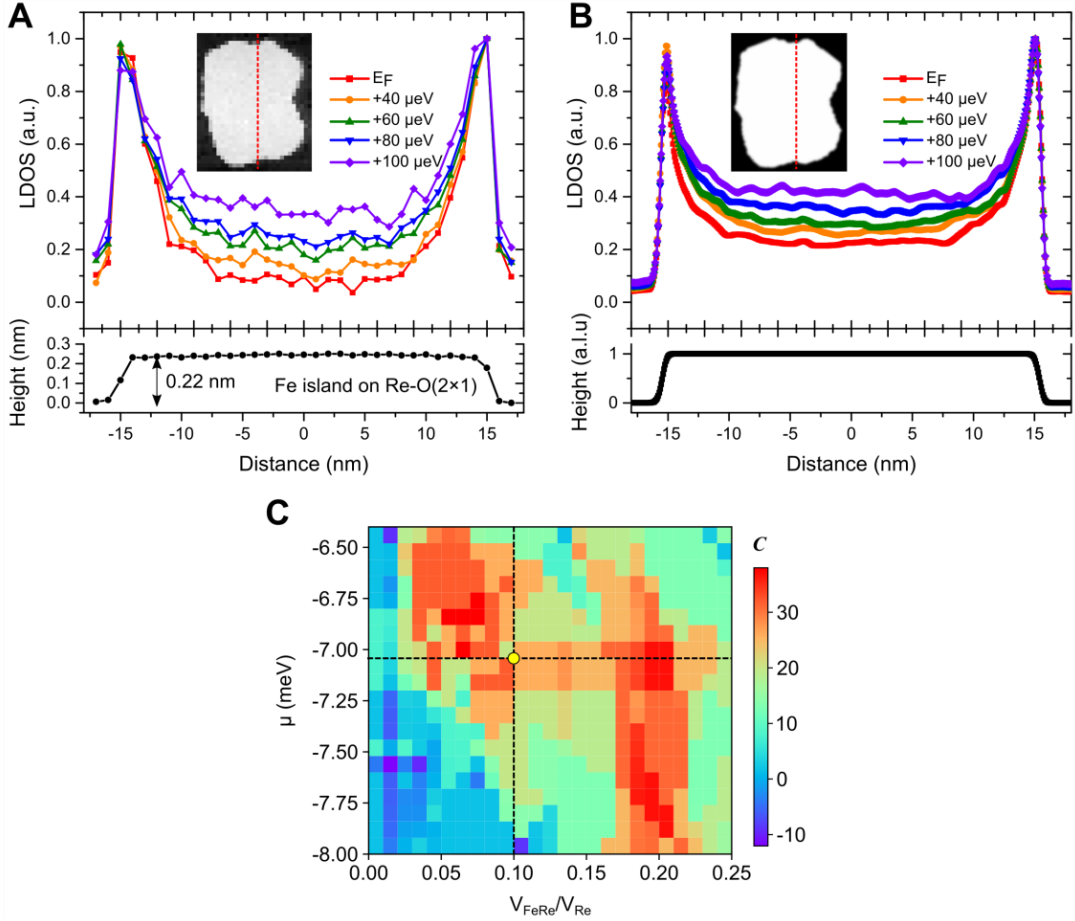

**Fig. 3. Decay of the chiral edge states inside an Fe island.** (**A**) Deconvoluted LDOS profiles obtained from the experimentally measured $dI/dU$ spectra along the red dotted line in the inset for several different energies: $E_F$, +40 μeV, +60 μeV, +80 μeV, and +100 μeV. (**B**) Theoretically computed LDOS along the red line in the inset for the same energies as in **A**. The corresponding surface profiles of the island are depicted in the lower panels of **A** and **B**. The theoretical results have been spatially convoluted to reproduce the experimental spatial resolution (see SI Sec. 3). Inset in **A** and **B**: the STM topography image and theoretically considered model structure of an Fe island, respectively. All LDOS profiles in **A** and **B** are normalized by their maximum values at the island's edge. (**C**) Theoretical phase diagram of the hybrid system. Chern number for an out-of-plane ferromagnetic structure of the Fe layer, as a function of $V_{FeRe}/V_{Re}$ and $\mu_{Fe}$ (see SI Sec. 3). The phase diagram reveals an abundance of topological phases near the parameter set used to describe the Fe/Re(0001)-O(2×1) hybrid system (yellow dot with crossed dotted lines).

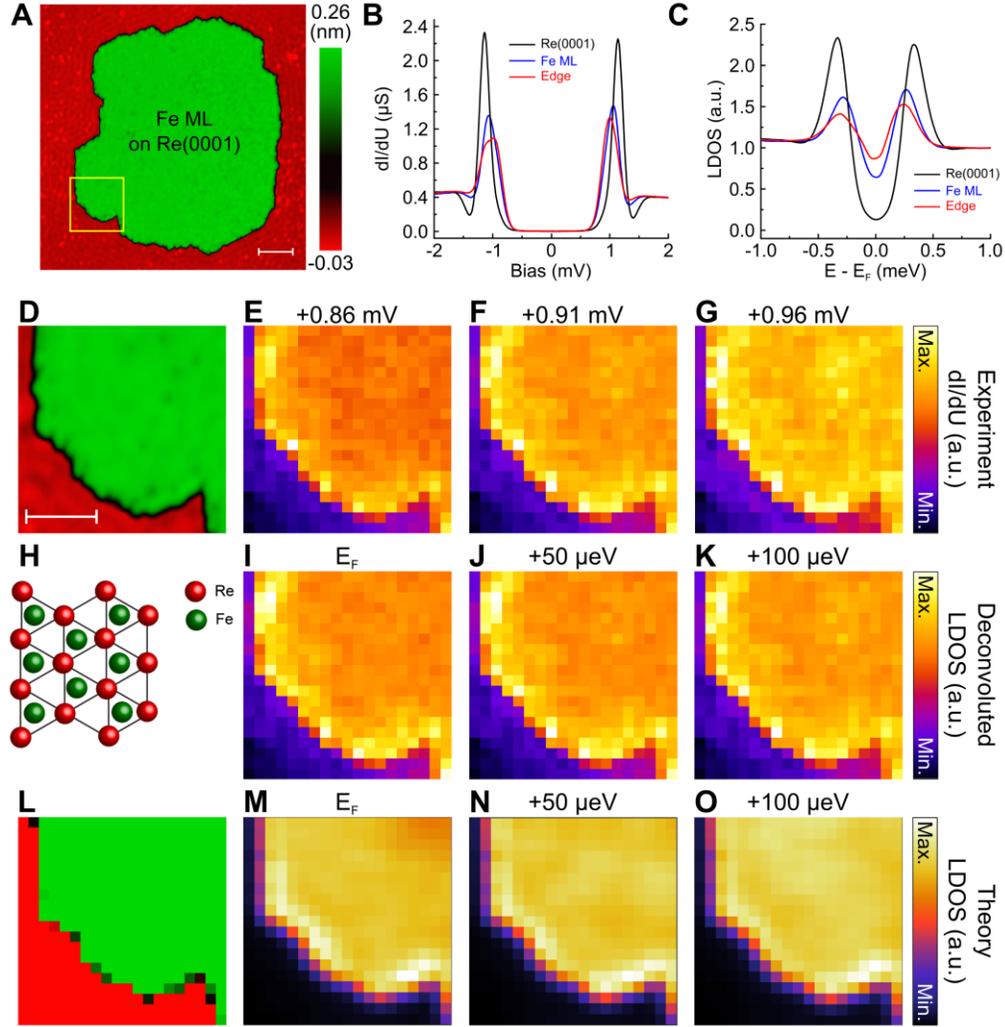

**Fig. 4. Topological trivial Fe island on a bare Re(0001) surface**. (**A**) The STM image of the Fe island being in direct contact with the superconducting Re substrate (scale bar: 10 nm). (**B**) Experimentally measured $dI/dU$ spectra on the clean Re(0001) surface (black), at the center (blue) and at the edge (red) of the Fe island using a superconducting Nb-tip whose gap size is $\Delta_{\text{tip}}^{\text{Nb}} = 0.86$ meV (see SI Sec. 2). (Tunneling conditions for STM/STS measurements: $U = 2.0$ mV, $I = 1.0$ nA, $T = 360$ mK). (**C**) The deconvoluted surface LDOS derived from the spectra shown in **B** taking into account the superconducting gap of the Nb tip used for the STS measurements. (see SI Sec. 2) (**D**) A zoomed-in STM image of the Fe island shown in **A** (yellow boxed area in **A**); (scale bar: 5 nm). (**E-G**) Experimentally measured $dI/dU$ maps for the Fe island shown in **D** at U=+0.86, +0.91 and +0.96 mV, which are corresponding to the energy of $E_F$, 50 and 100 µeV, respectively. (**H**) Atomic model of the Fe island on the bare Re(0001) substrate. Fe adatoms are located at the hollow sites of the Re substrate. (**I-K**) The spatial distribution of the deconvoluted surface LDOS for the island shown in **D**. (**L**) The zoomed-in image of the theoretically considered Fe island structure, reproducing the experimental Fe island in **D**. (**M-O**) Calculated LDOS of the island shown in **L** at the same energies as the experimental results. The theoretical data have been spatially convoluted to reproduce the experimental spatial resolution (see SI Sec. 3).

# Supplementary Materials for

**Atomic-Scale Interface Engineering of Majorana Edge Modes in a 2D Magnet-Superconductor Hybrid System**

Alexandra Palacio-Morales, Eric Mascot, Sagen Cocklin, Howon Kim, Stephan Rachel, Dirk K. Morr[*] and Roland Wiesendanger[*]

## Materials and Methods

### Experimental details and Data Acquisition

#### Preparation of the model-type hybrid sample and the STM-tip

An atomically clean Re(0001) surface was prepared in ultra-high vacuum (UHV) by repeated cycles of Ar$^+$-ion sputtering and annealing at temperatures between 1150°C and 1250°C followed by flashing the substrate at 1530°C [31]. The surface cleanliness was checked *in situ* by STM after transferring the sample into the cryostat. The ultrathin oxide layer was grown by flashing the Re single crystal in an O$_2$ atmosphere (the pressure range of $10^{-9} \sim 10^{-10}$ mbar) [32]. Fe was deposited by electron beam evaporation from a pure Fe rod (99.99+%) yielding a coverage of about 0.4 monolayers (ML).

The Nb-tip used for the STM/STS experiments was mechanically cut *ex situ* at ambient conditions and then inserted into the UHV system. It was conditioned by applying voltage pulses (few V for 50 ms) and gentle indentation into the Re(0001) surface. Its quality was judged acceptable if a sufficiently large Nb gap could be reproduced and no in-gap states were observed in the tunneling spectra obtained on the Re(0001) surface.

#### STM/STS data acquisition

The experiments were carried out in a $^3$He-cooled low-temperature STM system (USM-1300S, Unisoku, Japan) operating at T=360 mK under UHV conditions. Spectroscopic data were obtained by measuring the differential tunneling conductance (dI/dU) using a standard lock-in technique with a bias modulation voltage of 20 µV$_{rms}$ and a frequency of 841 Hz with opened feedback loop. The bias voltage was applied to the sample and the tunneling current was measured through the tip using a commercially available controller (Nanonis, SPECS, Swiss).

## Supplementary Text

### Section 1: Surface Characterization of Fe/Re(0001)-O(2×1)

A typical STM image of the Fe/Re(0001)-O(2×1) hybrid system with an individual Fe adatom (red dot) is shown in Fig.S1**A**. In general, it is not possible to identify the relative positions of the Fe- and O-atoms with respect to the Re surface atoms from the STM images. However, DFT studies [21] suggest that the O-atoms are located above the hcp hollow sites of the Re(0001) surface, forming a p(2×1) structure (see Fig.1**D** of the main text). This is confirmed by a surface profile (Fig.S1**B**, along the blue line in Fig.S1**A**) which reveals a periodicity of $2 \cdot \mathbf{a}_{Re} \cdot \cos(30°) = 0.47$ nm as expected for the oxygen-induced p(2×1) structure with the lattice constant of the Re(0001) surface, $\mathbf{a}_{Re} = 0.274$ nm.

The differential tunneling conductance, $dI/dU$, measured above the O(2×1) surface (Fig.S1**C,** blue line) reveals a coherence peak at $\Delta_{tip}^{Nb} + \Delta_{O(2\times1)}$, yielding a superconducting gap of $\Delta_{O(2\times1)} \approx$ 280 µeV in the O(2×1) layer which is slightly lower than that measured on the pure Re(0001) surface ($\Delta_{Re} \approx$ 330 µeV). Spectroscopic measurements on different structural domains of the O(2×1) layer indicate a uniform spatial distribution of the superconducting gap within our experimental resolution. The differential tunneling conductance (Fig. S1**C,** red line) measured near an isolated magnetic Fe adatom located on top of the O(2×1) surface (indicated by the red dot in Fig. S1**A**) reveals the existence of a Yu-Shiba-Rusinov (YSR) in-gap state at $E_{YSR} \approx \pm 145$ µeV (see arrows in Fig. S1**C**) arising from pair-breaking effects. This demonstrates that despite the presence of the intermediate oxide layer, the Fe adatoms still lead to magnetic scattering on the superconducting Re surface. However, the scattering strength is reduced by the intermediate oxide layer. When a single Fe adatom is placed directly above the hcp hollow site of the Re(0001) surface (without an intermediate oxide layer), the energy of the YSR state, $E_{YSR} \approx \pm 20$ µeV [12], is significantly reduced compared to that in the presence of the oxide layer, reflecting an increased interaction strength in the former case.

## Section 2: Deconvolution procedure of the tunneling spectra for the superconductor-superconductor tunneling junctions

In the STM configuration, the tunneling current ($I_T$) through the STM junction is proportional to the convolution of the tip density of states (TDOS, $\rho_T(E)$), surface density of states (SDOS, $\rho_S(E)$), and the Fermi-Dirac distribution function ($f(E,T)$). Therefore, the tunneling current and the differential tunneling conductance are expressed by

$$I(V,T) \propto \int \rho_S(E) \times \rho_T(E+eV) \times [f(E+eV,T) - f(E,T)] dE \quad \text{(S1)}$$

$$\frac{dI}{dV}(V,T) \propto \int \rho_S(E) \times \frac{\partial}{\partial V}\rho_T(E+eV) \times [f(E+eV,T) - f(E,T)] dE$$
$$+ \int \rho_S(E) \times \rho_T(E+eV) \times \left[\frac{\partial}{\partial V} f(E+eV,T)\right] dE \quad \text{(S2)}$$

where T is the temperature of the system.

Here, in this study, we were using a superconducting Nb tip to improve the energy resolution of the tunneling spectroscopy measurements. We assume a conventional BCS-type DOS for both the tip and the surface with the superconducting gaps of the tip and surface ($\Delta_T$, $\Delta_S$) and the broadening parameter $\Gamma$ [33] as given by

$$\rho_{T(S)}(E) = Re\left(\frac{E - i\Gamma}{\sqrt{(E-i\Gamma)^2 - \Delta_{T(S)}^2}}\right) \quad \text{(S3)}$$

We find $\Delta_T$, $\Delta_S$ and $\Gamma$ values of 1.41 (0.86) meV, 0.28 (0.30) meV and 0.023 (0.020) meV at the effective temperature, $T_{eff}$=0.4 K, by fitting the spectra obtained on the oxidized Re surface (the bare Re surface) away from the Fe island, and by considering the ac modulation of 20 µV$_{rms}$ used for the spectroscopic measurements (Fig. 1**E** and 4**B**, black). To extract the SDOS from the measured tunneling spectra using the superconducting Nb tip, it is necessary to deconvolute the SDOS from the measured spectra based on known information about the tip. As a mathematical approach for this procedure, we use a similar method as described in Ref. [34].

The Eq. (S2) is in the form of a Fredholm integral equation of the first kind:

$$\int_a^b K(x,y)h(x)dy = g(x) + \epsilon(x) \tag{S4}$$

where $g(x)$ is the measured spectrum with an error $\epsilon(x)$, $K(x,y)$ is the known function, which can be calculated with the assumed TDOS, $a$ and $b$ are finite integration limits and $h(x)$ is the unknown function to be found as the SDOS. This integral form can be rewritten in the discrete form of the linear equations [35]:

$$\sum_{i=0}^{n} A_{i,j} \cdot h_i = g_i + \epsilon_i \tag{S5}$$

The matrix $\boldsymbol{A}$ is defined by:

$$\boldsymbol{A}_{i,j} = K(x_j, y_i) \tag{S6}$$

Here, the matrix elements of $\mathbf{A}$ can be obtained by considering Eq. (S2):

$$A_{i,j} = \frac{\partial}{\partial V}\rho_T(E_j + eV_i)[f(E_j + eV_i, T) - f(E_j, T)] + \rho_T(E_j + eV_i)\frac{\partial}{\partial V}f(E_j + eV_i, T) \tag{S7}$$

To minimize the error, we assume $h(x)$ to be a smooth function and introduce an adjusting parameter γ as a prefactor of the trivial matrix $\boldsymbol{H}$, which can be arbitrarily chosen [36].
Finally, we can obtain the column vector of the deconvoluted spectrum by applying the operation into the column vector of the measured spectrum as given by:

$$(\boldsymbol{A}^*\boldsymbol{A} + \gamma \boldsymbol{H})^{-1}\boldsymbol{A}^* \cdot \left[\overrightarrow{\frac{dI}{dV}}\right]_{measured} \sim [\overrightarrow{\rho_S}] \tag{S8}$$

All deconvoluted spectra (Fig. 1**F** and 4**C**) and differential tunneling conductance maps (Fig. 2**G**-**L** and Fig. 4**I**-**K**) in the main text are computed by applying the method described above. Note that we calculate the matrix **A** whenever the measurements were performed with different tips because the superconducting gap of the Nb tip strongly depends on the condition of the tip apex, i.e. its crystalline structure and shape [37].

**Section 3: Theoretical Model**

To investigate the electronic structure of the Fe/Re(0001)-O(2×1) hybrid systems, we employ a Slater-Koster [38] tight-binding Hamiltonian which models the Re bulk system with the relevant surface structure as a two-dimensional system. The resulting Hamiltonian is given by:

$$H = H_{Fe} + H_{Re} + H_{Fe-Re} \tag{S9}$$

$$H_{Fe} = \sum_{\langle \mathbf{r}_1,\mathbf{r}_2 \rangle} \left( d^\dagger_{\mathbf{r}_1} \tau_{Fe}(\mathbf{r}_1 - \mathbf{r}_2) d_{\mathbf{r}_2} + h.c. \right) + \sum_{\mathbf{r}} d^\dagger_{\mathbf{r}} \xi_{Fe}(\mathbf{r}) d_{\mathbf{r}} \tag{S10}$$

$$H_{Re} = \sum_{\langle \mathbf{r}_1,\mathbf{r}_2 \rangle} \left( c^\dagger_{\mathbf{r}_1} \tau_{Re}(\mathbf{r}_1 - \mathbf{r}_2) c_{\mathbf{r}_2} + h.c. \right) + \sum_{\mathbf{r}} c^\dagger_{\mathbf{r}} \xi_{Re}(\mathbf{r}) c_{\mathbf{r}} \tag{S11}$$

$$H_{FeRe} = \sum_{\langle \mathbf{r}_1,\mathbf{r}_2 \rangle} \left( d^\dagger_{\mathbf{r}_1} \tau_{FeRe}(\mathbf{r}_1 - \mathbf{r}_2) c_{\mathbf{r}_2} + h.c. \right) \tag{S12}$$

where the spinors are defined via

$$c^\dagger_{\mathbf{r}} = \left( c^\dagger_{\mathbf{r},\uparrow}, c^\dagger_{\mathbf{r},\downarrow}, c_{\mathbf{r},\downarrow}, -c_{\mathbf{r},\uparrow} \right) \tag{S13}$$

$$c^\dagger_{\mathbf{r},\sigma} = \left( c^\dagger_{\mathbf{r},\sigma,d_{xy}}, c^\dagger_{\mathbf{r},\sigma,d_{yz}}, c^\dagger_{\mathbf{r},\sigma,d_{zx}}, c^\dagger_{\mathbf{r},\sigma,d_{x^2-y^2}}, c^\dagger_{\mathbf{r},\sigma,d_{3r^2-z^2}} \right) \tag{S14}$$

and similarly for $d^\dagger_{\mathbf{r}}$. The electron creation operators $d^\dagger_{\mathbf{r}}$ ($c^\dagger_{\mathbf{r}}$) create an electron in the Fe 3d orbitals (Re 5d orbitals) at site $\mathbf{r}$. $\tau$ and $\xi$ describe the hopping and on-site interaction parameters, respectively. The hopping interactions are given by:

$$\tau_{Fe}(\mathbf{r}_1 - \mathbf{r}_2) = \tau_z \sigma_0 \sum_{\beta=\{\sigma,\pi,\delta\}} V^{dd\beta}_{Fe} A^{dd\beta}(\boldsymbol{\delta}_{12}) + \alpha_{Fe} \tau_z (\boldsymbol{\sigma} \times \boldsymbol{\delta}_{12})_z L_0 \tag{S15}$$

$$\tau_{Re}(\mathbf{r}_1 - \mathbf{r}_2) = \tau_z \sigma_0 \sum_{\beta=\{\sigma,\pi,\delta\}} V^{dd\beta}_{Re} A^{dd\beta}(\boldsymbol{\delta}_{12}) + \alpha_{Re} \tau_z (\boldsymbol{\sigma} \times \boldsymbol{\delta}_{12})_z L_0 \tag{S16}$$

$$\tau_{FeRe}(\mathbf{r}_1 - \mathbf{r}_2) = \tau_z \sigma_0 \sum_{\beta=\{\sigma,\pi,\delta\}} V^{dd\beta}_{FeRe} A^{dd\beta}(\boldsymbol{\delta}_{12}) \tag{S17}$$

$$\boldsymbol{\delta}_{12} = \frac{\mathbf{r}_1 - \mathbf{r}_2}{|\mathbf{r}_1 - \mathbf{r}_2|} \tag{S18}$$

where $\boldsymbol{\tau}$ and $\boldsymbol{\sigma}$ are Pauli matrices acting on particle-hole and spin degrees of freedom respectively, $A^{dd\beta}$ are the Slater Koster matrices for $\sigma$, $\pi$, and $\delta$ bonds between d orbitals, $V^{dd\beta}$ are the tight-binding hopping integrals, and $\alpha$ is the Rashba spin-orbit coupling. For the results shown in the main text, the lattice structure of the above Hamiltonian (which determined the form of $A^{dd\beta}$) is shown in Fig. 1**D** of the main text. The on-site interactions are given by:

$$\xi_{Fe}(\mathbf{r}) = \tau_z(\mu_{Fe}\sigma_0 L_0 + \lambda_{Fe}\mathbf{S}\cdot\mathbf{L}) + \tau_x \sigma_0 \Delta_{Fe} + J\tau_0 \frac{\sigma_z}{2} L_0 \tag{S19}$$

$$\xi_{Re}(\mathbf{r}) = \tau_z(\mu_{Re}\sigma_0 L_0 + \lambda_{Re}\mathbf{S}\cdot\mathbf{L}) + \tau_x \sigma_0 \Delta_{Re} \tag{S20}$$

where $\mu$ is the chemical potential, $\lambda$ is the spin-orbit coupling, $\Delta$ is the superconducting s-wave order parameter, and J is the magnetic exchange coupling. Explicitly, the spin-orbit term is given by:

$$\mathbf{S} \cdot \mathbf{L} = \frac{1}{2} \sum_i \sigma_i L_i = \frac{1}{2} \begin{pmatrix} L_z & L_x - iL_y \\ L_x + iL_y & -L_z \end{pmatrix} \tag{S21}$$

$$L_x = \begin{pmatrix} 0 & 0 & -i & 0 & 0 \\ 0 & 0 & 0 & -i & -\sqrt{3}i \\ i & 0 & 0 & 0 & 0 \\ 0 & i & 0 & 0 & 0 \\ 0 & \sqrt{3}i & 0 & 0 & 0 \end{pmatrix} \tag{S22}$$

$$L_y = \begin{pmatrix} 0 & i & 0 & 0 & 0 \\ -i & 0 & 0 & 0 & 0 \\ 0 & 0 & 0 & -i & \sqrt{3}i \\ 0 & 0 & i & 0 & 0 \\ 0 & 0 & -\sqrt{3}i & 0 & 0 \end{pmatrix} \tag{S23}$$

$$L_z = \begin{pmatrix} 0 & 0 & 0 & 2i & 0 \\ 0 & 0 & i & 0 & 0 \\ 0 & -i & 0 & 0 & 0 \\ -2i & 0 & 0 & 0 & 0 \\ 0 & 0 & 0 & 0 & 0 \end{pmatrix} \tag{S24}$$

We use a constant pairing term between orbitals of opposite $L_z$, i.e. $\Delta_0^{Fe(Re)} c_{m,\uparrow}^\dagger c_{-m,\downarrow}^\dagger + h.c.$ which then yields for the superconducting order parameter in the cubic harmonic basis:

$$\Delta_{Fe(Re)} = \Delta_0^{Fe(Re)} \begin{pmatrix} 1 & 0 & 0 & 0 & 0 \\ 0 & -1 & 0 & 0 & 0 \\ 0 & 0 & -1 & 0 & 0 \\ 0 & 0 & 0 & 1 & 0 \\ 0 & 0 & 0 & 0 & 1 \end{pmatrix} \tag{S25}$$

where the sign difference between the order parameters in the $\Delta_{xz}$ and $\Delta_{yz}$ on one hand, and $\Delta_{xy}$, $\Delta_{x^2-y^2}$ and $\Delta_{3r^2-z^2}$ is due to $Y_l^m = (-1)^m Y_l^{-m}$. We fix the superconducting order parameter of iron and rhenium, such that the energies of the coherence peaks match the experimentally measured ones, and scale down the band parameters for iron [39] to 0.8% and rhenium [39] to 10% of the original theoretical values such that the coherence length also matches the experimental values. We use the hopping integrals from rhenium scaled by 1/10 for the hopping integrals between iron and rhenium to simulate the effect of the oxide layer. The parameters are listed in Table S1. Note that Fe itself is not superconducting; superconductivity is only proximity-induced from the Re substrate. In order to describe the magnet-superconductor hybrid system efficiently, we consider only the effective two-dimensional system, as mentioned before. As a consequence, a superconducting order parameter for Fe, $\Delta_{Fe}$, explicitly appears in the Hamiltonian, to account for the proximity-induced superconductivity.

## Calculation of Local Density of States

The orbital and spin resolved LDOS at site *r*, orbital α, spin σ, and energy E is obtained from the retarded Green's function via

$$\rho(r, \alpha, \sigma, E = \hbar\omega) = -\frac{1}{\pi} \text{Im} G_{ret}(r, \alpha, \sigma, r, \alpha, \sigma, \omega) \tag{S26}$$

where the retarded Green's function is given by

$$G_{ret}(r, \alpha, \sigma, r', \alpha', \sigma', \omega) = \langle r, \alpha, \sigma | (\omega + i\delta - H)^{-1} | r', \alpha', \sigma' \rangle \\ = \sum_k \frac{\langle r, \alpha, \sigma | k \rangle \langle k | r', \alpha', \sigma' \rangle}{\omega + i\delta - E_k} \tag{S27}$$

and *k* and $E_k$ are the eigenvectors and eigenvalues of the Hamiltonian matrix *H*. The local density of states at site *r* and energy E are the sum of the orbital and spin resolved LDOS at site *r* and energy E.

$$\rho(r, E) = \sum_{\alpha, \sigma} \rho(r, \alpha, \sigma, E) \tag{S28}$$

Only states with $E_k \approx \omega$ contribute to the sum, and we therefore calculated 100 eigenvectors at each energy using the Lanczos algorithm. The spatial LDOS was calculated for each site and smoothed using a Gaussian kernel convolution to simulate the STM tip interacting with nearby sites. To match the spatial resolution of the experiment, the spatial LDOS was pixelated where the pixel width, $w_p$, is the spatial resolution and the LDOS for sites inside a pixel were summed. We used $w_p = 1$ nm, and $w_p = 0.33$ nm for Figs. 2 and 4 of the main text, respectively, and $w_p = 0.7$ nm for Fig. S5.
In Fig. S2, we present the resulting LDOS on the Re(0001)-O(2×1) surface, and on the Fe ML, which reproduce the experimentally measured positions of the coherence peaks.

## Calculation of the Chern number

Let *P(k)* denote the projector onto occupied states in crystal momentum space

$$P(k) = \sum_n |n(k)\rangle\langle n(k)| \, \Theta(-E_{n(k)}) \tag{S29}$$

where *n(k)* is the band index at crystal wave vector *k* and $\Theta(-E_{n(k)})$ is the Heaviside function at energy $-E_{n(k)}$. The Chern number is given by the formula:

$$C = \frac{1}{2\pi i} \int_{BZ} \text{Tr}\left(P(k)\left[\partial_{k_x} P(k), \partial_{k_y} P(k)\right]\right) dk_x dk_y \tag{S30}$$

where Tr denotes the trace, square brackets denote the commutator, and the integral runs over the full Brillouin zone. The phase diagram in Fig. 3**C** of the main text was generated by computing $P(\mathbf{k})$ at points in the neighborhood of $\mathbf{k}$ to approximate the partial derivatives using the central difference method [40], then integrated using an adaptive quadrature rule [41]. The parameter $V_{FeRe}/V_{Re}$ used in Fig. 3**C** is the ratio of hybrid coupling to rhenium coupling which is kept the same for each bond, i.e. $V_{FeRe}^{dd\alpha}/V_{Re}^{dd\alpha} = V_{FeRe}/V_{Re}$ for $\alpha \in \{\sigma, \pi, \delta\}$.

**Section 4: Magnetic Structure of the Fe Island and Topological Superconductivity**

We next consider a 120° Néel-ordered in-plane structure of the Fe moments [29], as shown in Fig. S3**A**. In Fig. S3**B**, we show the resulting superconducting gap and Chern number as a function of $\mu_{Fe}$. The chemical potential used in the main text is indicated by a green arrow. Similarly to the ferromagnetic, out-of-plane structure discussed in the main text, we find that the system is in a topological phase with Chern number $C = 3$.

For a magnetically disordered Fe island (i.e., for a completely random orientation of the magnetic moments) the hybrid system is gapless, and hence in a trivial, non-topological phase. As a result, the LDOS at $E_F$ shows delocalized, bulk-like excitations in the Fe island, as shown in Fig. S4.

**Section 5: Evolution of Topological Majorana modes in a Generic Topological Superconductor**

To demonstrate the universality of the combined spatial and energy evolution of Majorana edge modes shown in Fig. 2 of the main text, we consider a previously introduced generic model for a topological superconductor described by the Hamiltonian [15,17]

$$H = \sum_{r,\delta} c_r^\dagger \tau_z(-t\sigma_0 + i\alpha(\boldsymbol{\sigma} \times \boldsymbol{\delta}))c_{r+\delta} + \sum_r c_r^\dagger(-\mu\tau_z\sigma_0 + \Delta_0\tau_x\sigma_0)c_r + \sum_R c_R^\dagger J\tau_0\sigma_z c_R \quad (S31)$$

where

$$c_r^\dagger = (c_{r\uparrow}^\dagger, c_{r\downarrow}^\dagger, c_{r\downarrow}, -c_{r\uparrow}) \quad (S32)$$

and **r** runs over all sites, **δ** runs over hopping directions, **R** runs over magnetic impurity sites, **τ** and **σ** are Pauli matrices acting on particle-hole and spin degrees of freedom, respectively, t is the hopping integral, α is the Rashba spin-orbit coupling, μ is the chemical potential, $\Delta_0$ is the superconducting s-wave order parameter, and J is the magnetic exchange coupling. We consider a two-dimensional square lattice, with a circular magnetic island of radius $15a_0$, as shown in Fig. S5**A** In Fig. S5**B-G**, we present the energy evolution of the LDOS, which is qualitatively similar to that shown in Fig. 2 of the main text for the Fe/Re(0001)-O(2×1) hybrid structure. Moreover, a plot of the LDOS as a function of position (along a cut through the island) and energy shown in Fig. S5**H** reveals that the chiral Majorana mode is located at the edge of the island for all energies up to the superconducting gap edge.

We therefore conclude that the combined spatial and energy evolution of the topological Majorana edge modes, as reflected in the LDOS, is universal, and does not depend on a particular band structure.

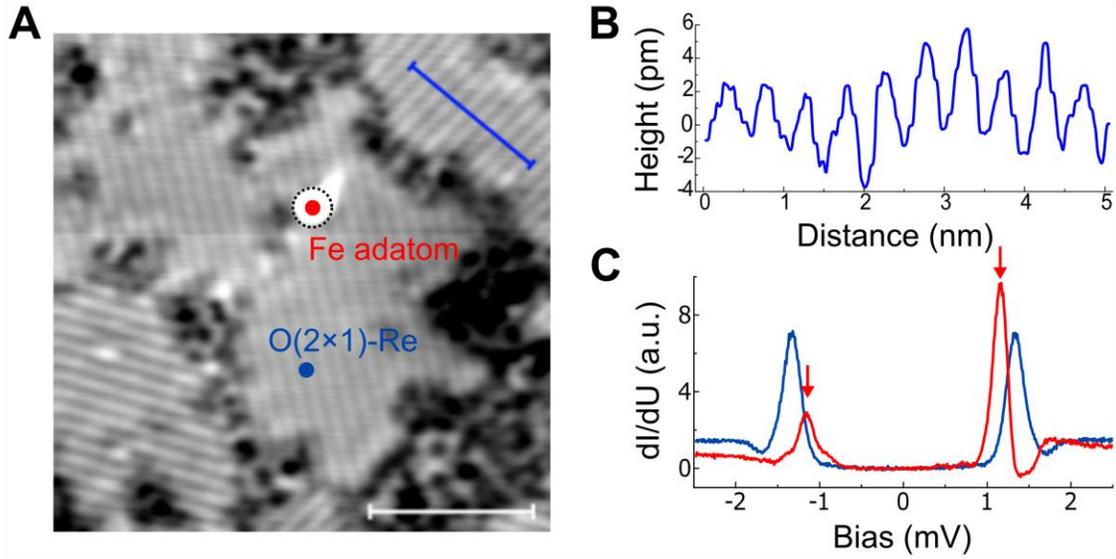

**Figure S1. STM/STS characterization of the Re(0001)-O(2×1) surface.** (**A**) STM topographic image revealing several structural domains of the oxygen-induced (2×1) reconstruction of Re(0001). (**B**) Height profile taken along the blue line in S1**A**. (**C**) The differential tunneling conductance (dI/dU), measured above the O(2×1) surface (blue), reveals coherence peaks at $\pm|\Delta_{tip}^{Nb} + \Delta_{O(2\times1)}|$, yielding a superconducting gap of $\Delta_{O(2\times1)} \approx 280$ μeV which is slightly lower than that measured on the bare Re(0001) surface ($\Delta_{Re} \approx 330\ \mu eV$). The spectrum obtained above a magnetic Fe adatom (red) shows YSR bound states (red arrows) at $\pm|\Delta_{tip}^{Nb} + E_{YSR}|$, $\Delta_{tip}^{Nb} =$ 1.04 meV and $E_{YSR} = 145$ μeV. Measurement conditions: U=2.5 mV, I=1.0 nA, T= 360 mK.

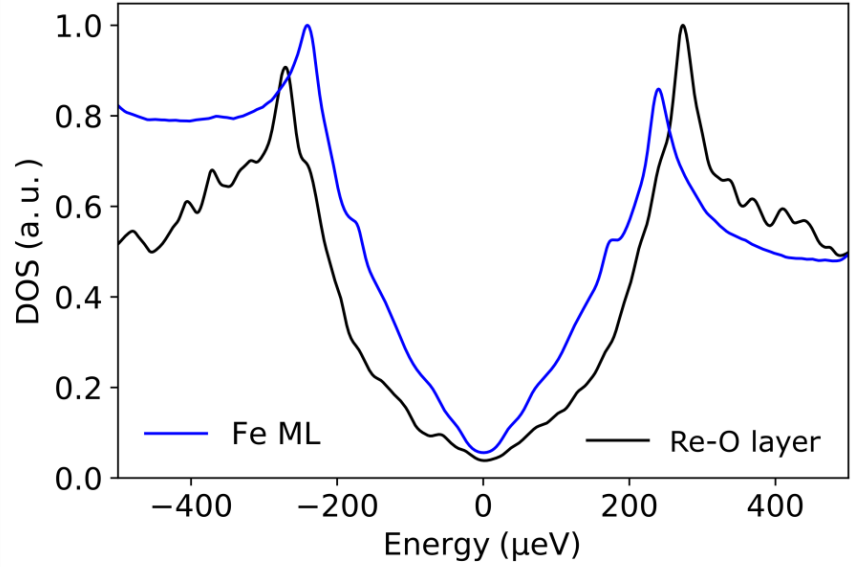

**Figure S2. Theoretically computed LDOS.** Theoretically computed LDOS on the Re(0001)-O(2×1) surface (black line) and on the Fe ML (blue line). The position of the coherence peaks is shifted from $\Delta_{O(2\times1)} \approx 280~\mu eV$ on the Re(0001)-O(2×1) surface to $\Delta_{Fe} \approx 240~\mu eV$ on the Fe ML. The finite LDOS at zero energy arises from the finite electronic damping, $\Gamma$, introduced for computational reasons in the calculations.

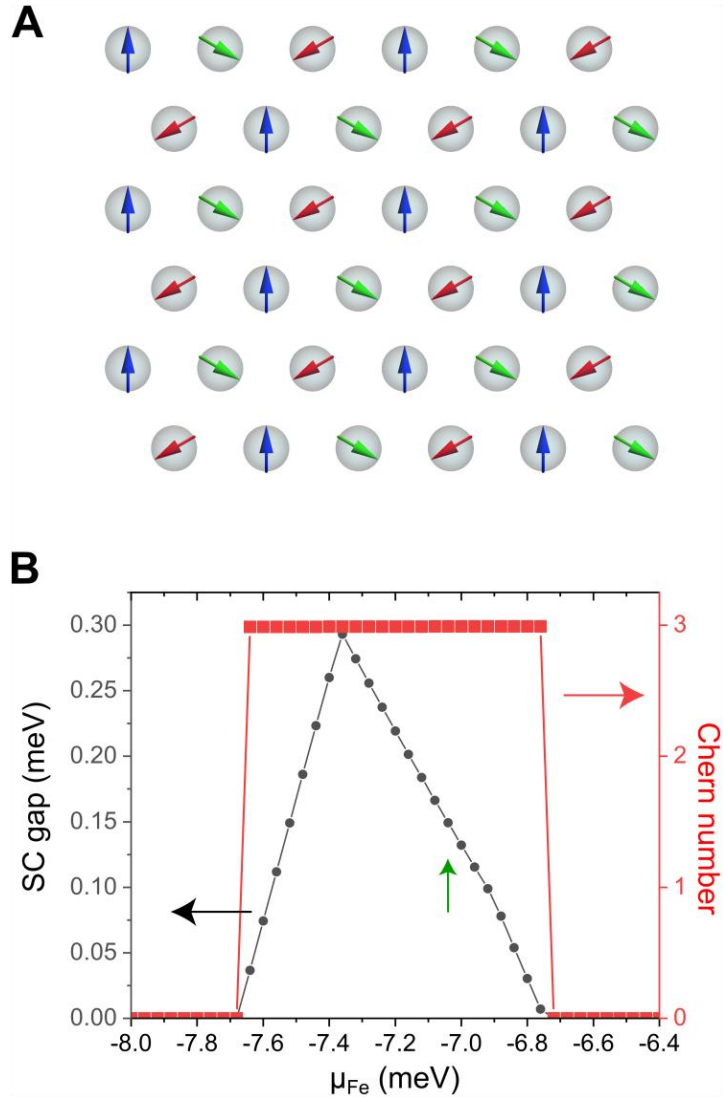

**Figure S3. Topological phase diagram.** (**A**) 120° Néel-ordered in-plane configuration of the Fe moments. (**B**) Superconducting gap (black) and Chern number (red) for a range of $\mu_{Fe}$. The Chern number is 3, when the system is gapped, and 0 when the gap closes. The hybrid couplings $V_{FeRe}$ are set to zero, $\alpha_{Fe}$=1.6 meV, $\lambda_{Fe}$=0.48 meV, and $\Delta_{Fe}$=2.24 meV. As expected, the Chern number changes when the gap closes.

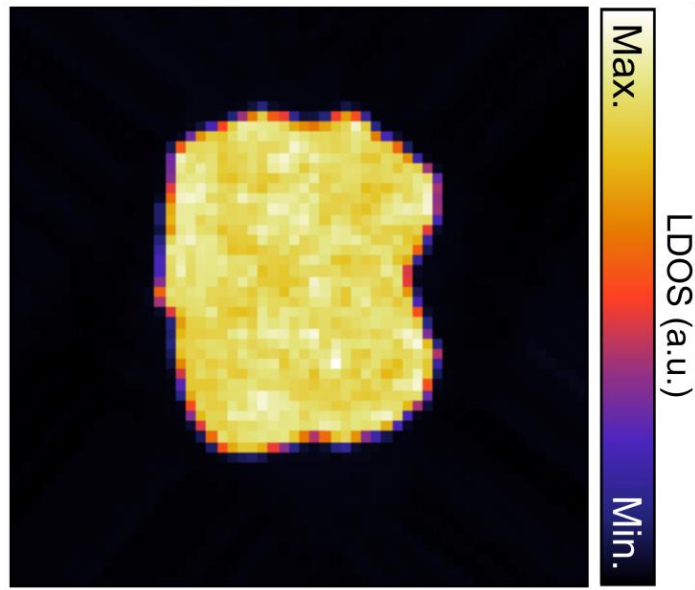

**Figure S4. Spatially resolved LDOS at $E_F$ of a magnetically disordered Fe island.** Magnetic Fe moments are randomized over a unit sphere. Parameters from Table S1 were used.

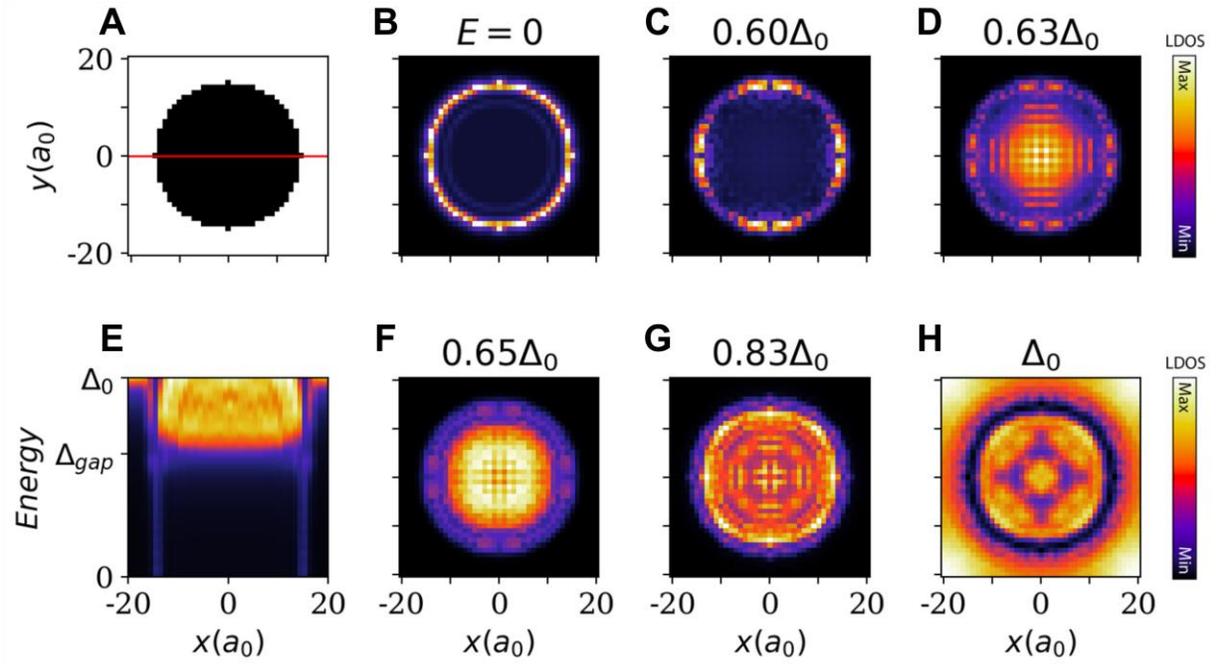

**Figure S5. Energy evolution of the spatially resolved LDOS for a generic topological superconductor.** (**A**) Schematic picture of a ferromagnetic Shiba island with large black circles denoting magnetic defect sites. (**B-G**) Spatially resolved LDOS from $E_F$ to $\Delta_0$ for $\mu=-4t$, $\alpha=0.8t$, $\Delta_0=1.2t$, and $J=2t$. Here, $\Delta_0$ is the s-wave order parameter and $\Delta_{gap}$ is the calculated gap in the magnetic island ($\Delta_{gap}=0.62\Delta_0$). Such a large value of $\Delta_0$ is necessary in order to obtain a coherence length that is smaller than the size of the island, otherwise no well-defined edge modes can be observed. As the energy increases, the weight of the LDOS moves from the edge of the magnetic island toward the center, similar to the results shown in Fig.2 of the main text. (**H**) Intensity plot of the LDOS as a function of position (along a cut through the island) and energy, showing the energy evolution of the edge modes.

|  | Fe | Re | Fe-Re |
|---|---|---|---|
| $V^{dd\sigma}$ | -3.8561 meV | -144.19 meV | -14.419 meV |
| $V^{dd\pi}$ | 2.1134 meV | 66.763 meV | 6.6763 meV |
| $V^{dd\delta}$ | -0.19264 meV | -7.4015 meV | -0.74015 meV |
| $\mu$ | -7.04 meV | -71.5 meV |  |
| $\lambda$ | 2.4 meV | 32 meV |  |
| $\Delta_0$ | 0.45 meV | 0.28 meV |  |
| $J$ | 17.473 meV |  |  |
| $\alpha$ | 2.4 meV |  |  |

**Table S1. Tight binding parameters for the Fe/Re(0001)-O(2×1) hybrid system.**